\documentclass[a4, 10pt]{article}
\title{Validity and failure of  some entropy inequalities 
for  CAR systems}
\date{}
\author{Hajime Moriya} 
\usepackage{latexsym}
\usepackage{amssymb} 
\usepackage{amsmath} %
\usepackage{amsthm} %
\usepackage{amscd} %
\usepackage{overcite}
\makeatletter
\def\@cite#1{\textsuperscript{#1}} 
\makeatother
\makeatletter
\def\@biblabel#1{#1}
\makeatother
\newcommand{\LAT}{{\mathcal {L}}}%
\newcommand{\qedb}{\hbox{\rule[-2pt]{3pt}{6pt}}}%
\newcommand{\I}{{\mathrm{I}}}%
\newcommand{\J}{{\mathrm{J}}}%
\newcommand{\K}{{\mathrm{K}}}%
\newcommand{\Bl}{{\cal B}}%
\newcommand{\IaJ}{{\I}\cap {\J}}%
\newcommand{\IuJ}{{\I}\cup {\J}}%
\newcommand{\KuI}{{\K}\cup {\I}}%
\newcommand{\KuJ}{{\K}\cup {\J}}%
\newcommand{\IuK}{{\I}\cup {\K}}%
\newcommand{\Al}{{\cal{A}}}%
\newcommand{\Almath}{{\mathfrak{A}}}%
\newcommand{\Ali}{\Al_{i}}%
\newcommand{\AlI}{{\cal A}({\I})}%
\newcommand{\AlJ}{{\cal A}({\J})}%
\newcommand{\AlK}{{\cal A}({\K})}%
\newcommand{\AlJp}{{\cal A}(\J)_{+}}%
\newcommand{\AlJm}{{\cal A}(\J)_{-}}%
\newcommand{\AlIp}{{\cal A}(\I)_{+}}%
\newcommand{\AlIm}{{\cal A}(\I)_{-}}%
\newcommand{\AlIpm}{{\cal A}(\I)_{\pm}}%
\newcommand{\AlKp}{{\cal A}(\K)_{+}}%
\newcommand{\AlIuJp}{{\cal A}(\IuJ)_{+}}%
\newcommand{\AlKuI}{{\cal A}(\KuI)}%
\newcommand{\AlKuJ}{{\cal A}(\KuJ)}%
\newcommand{\AlIaJ}{{\cal A}({\I}\cap {\J})}%
\newcommand{\AlIuJ}{{\cal A}({\I}\cup {\J})}%
\newcommand{\AlKIJ}{{\cal A}({\K}\cup {\I} \cup  {\J})}%
\newcommand{\AlIcommut}{\AlI^{\prime}}%
\newcommand{\AlIpJ}{\Al(\I^{\prime}\,\vert\, \J)}%
\newcommand{\AlIpIJ}{\Al(\I^{\prime}\,\vert\, \IuJ)}%
\newcommand{\vp}{\varphi}
\newcommand{\lam}{\lambda}%
\newcommand{\ome}{\omega}
\newcommand{\omef}{\ome_{1}}
\newcommand{\omes}{\ome_{2}}
\newcommand{\vrho}{\varrho}%
\newcommand{\vrhof}{\vrho_{1}}
\newcommand{\vrhoft}{\vrho_{1}\Theta}
\newcommand{\vrhofb}{\overline{\vrhof}}%
\newcommand{\vrhos}{\vrho_{2}}
\newcommand{\vrhostil}{\widetilde{\vrho_{2}}}
\newcommand{\vrhoI}{\vrho_{\I}}
\newcommand{\vrhoJ}{\vrho_{\J}}
\newcommand{\vrhoK}{\vrho_{\K}}
\newcommand{\vrhoKI}{\vrho_{\KuI}}
\newcommand{\vrhoKJ}{\vrho_{\KuJ}}
\newcommand{\vrhoKt}{\vrho_{\K}\Theta}
\newcommand{\pivrhof}{\pi_{\vrhof}}%
\newcommand{\pivrhoft}{\pi_{\vrhoft}}%
\newcommand{\pivrhoK}{\pi_{\vrhoK}}%
\newcommand{\pivrhoKt}{\pi_{\vrhoKt}}%
\newcommand{\pivrhostil}{\pi_{\vrhostil}}%
\newcommand{\vrhofo}{\overline{\vrhof}}%
\newcommand{\vpI}{\vp_{\I}}%
\newcommand{\vpJ}{\vp_{\J}}%
\newcommand{\vpIuJ}{\vp_{\IuJ}}%
\newcommand{\vpKI}{\vp_{\K\cup \I}  }%
\newcommand{\vpKJ}{\vp_{\K\cup \J}}%
\newcommand{\aicr}{a_i^{\ast}}%
\newcommand{\ai}{a_i}%
\newcommand{\ajcr}{a_j^{\ast}}%
\newcommand{\aj}{a_j}%
\newcommand{\identitybf}{{\mathbf{1} } }
\newcommand{\id}{{\mathbf{1} } }
\newcommand{\proofend}{{\hfill \qedb}}
\newcommand{\cstar}{{\bf C}^{\ast}}%
\newcommand{\nonum}{\nonumber}%
\newcommand{\Tr}{\mathbf{Tr}}%
\newcommand{\EJ}{E_{\J}}%
\newcommand{\EI}{E_{\I}}%
\newcommand{\EIaJ}{E_{\I \cap \J} }%
\newcommand{\vpi}{\vp_{i}}%
\newcommand{\Hil}{{\cal H}}%
\newcommand{\Kil}{{\cal K}}%
\newcommand{\Hilf}{\Hil_{1}}%
\newcommand{\Hils}{\Hil_{2}}%
\newcommand{\Hilfs}{\Hil_{1,2}}%
\newcommand{\Ap}{A_{+}}%
\newcommand{\Am}{A_{-}}%
\newcommand{\Apm}{A_{\pm}}%
\newcommand{\Af}{A_{1}}%
\newcommand{\As}{A_{2}}%
\newcommand{\Asp}{A_{2+} }%
\newcommand{\Asm}{A_{2-} }%
\newcommand{\Aspm}{A_{2\pm} }%
\newcommand{\Ome}{\mit\Omega}%
\renewcommand{\Vec}{\Ome}%
\newcommand{\Vecvrhof}{\Ome_{\vrhof}}%
\newcommand{\Vecvrhostil}{\Ome_{\vrhostil}}%
\newcommand{\Hilvrhof}{{\cal H}_{\vrhof}}%
\newcommand{\Hilvrhostil}{{\cal H}_{\vrhostil}}%
\newcommand{\Hilome}{{\cal H}_{\ome}}%
\newcommand{\piome}{\pi_{\ome}}%
\newcommand{\Vecome}{\Ome_{\ome}}%
\newcommand{\vi}{v_{i}}%
\newcommand{\vI}{v_{\I}}%
\newcommand{\vIuJ}{v_{\IuJ}}%
\newcommand{\vJ}{v_{\J}}%
\newcommand{\vK}{v_{\K}}%
\newcommand{\phivec}{{\mit \Phi}}%
\newcommand{\psivec}{{\mit \Psi}}%
\newcommand{\BHvrhof}{\Bl(\Hilvrhof)}
\newcommand{\uone}{u_{1}}%
\begin{document}
\maketitle
\theoremstyle{plain}%
\newtheorem{lem}{Lemma}
\newtheorem{pro}[lem]{Proposition}%
\newtheorem{thm}[lem]{Theorem}
\ \\
\ \\
\ \\
\ \\
{\large{
\begin{abstract}
Basic properties of von Neumann entropy 
such as  the triangle inequality and what we call MONO-SSA
are studied for CAR systems.
We show that   both  inequalities
 hold   for every  even state
 by   using  symmetric purification  which is 
 applicable to such a state.
We construct a certain class of 
 noneven states giving  examples 
of  the nonvalidity of  those  inequalities.  
\end{abstract}
}}
%
%
%
\newpage
\section{Introduction}
\label{sec:INTRO}
Let $\Hil$ be a  Hilbert space and 
$D$ be a density matrix on $\Hil$,
 i.e. a positive trace class operator on $\Hil$
 whose trace is unity.
The   von Neumann entropy
 is given by
\begin{eqnarray}
\label{eq:vn}
-\Tr\bigl( D \log D  \bigr),\
\end{eqnarray}
where $\Tr$ denotes   the  trace  which takes the value $1$ 
 on each minimal projection.
 Let  $\vrho$ be  a normal state of $\Bl(\Hil)$,
 the  set of all bounded linear operators on $\Hil$.
 Then   $\vrho$ has its  
 density matrix  $D_{\vrho}$, and
its   von Neumann  entropy
 $S(\vrho)$ is given  by    (\ref{eq:vn})
 with $D=D_{\vrho}$.

It has been  known that     
 von Neumann entropy is useful   for description and 
 characterization
  of   state correlation   for    composite systems.
Among others, the  following inequality  called 
strong subadditivity (SSA) is remarkable:
\begin{eqnarray}
\label{eq:SSA}
S(\vp_{\I\cup \J})-S(\vp_\I)-S(\vp_{\J})+S(\vp_{\I\cap \J}) \leq 0,
\end{eqnarray}
 where $\I$, $\J$,  
$\IaJ$ and $\IuJ$
denote the  indexes of  subsystems 
  and $\vpI$  denotes the 
 restriction of a  state  $\vp$ 
 to the subsystem indexed by  $\I$, and so on.
  Such  entropy inequalities 
have been studied  for quantum systems,  
see e.g. Refs. \citen{BRA2}, \citen{73lett}, 
\citen{OP}, 
\citen{MBR}, 
 \citen{THIR}, \citen{AW}, 
and also their references.
However, the composite systems considered there 
were   mostly      tensor product  of 
 matrix algebras to which we  refer as the 
 tensor product systems.

We   investigate  
some well known entropy inequalities, the triangle 
 inequality and MONO-SSA (which will be specified soon),   
for CAR systems.
This study   is   relevant  to  our previous  works  on  
   state correlations such as quantum-entanglement  \cite{ENTA} 
and separability \cite{SEP} for CAR systems. 
In a certain sense, the   conditions  of 
validity and failure of  such entropy inequalities
which we are going to establish
will explain   the   similarities  and differences
 in  the possible forms of state correlations  
 between   CAR and tensor product  systems.

Let $\LAT$ be  an arbitrary  discrete set.
The canonical anticommutation relations (CAR)  are 
\begin{eqnarray*}
\{ \aicr, \aj \}&=&\delta_{i,j}\, \identitybf,  \\
\{ \aicr, \ajcr \}&=&\{ \ai, \aj \}=0,
\end{eqnarray*} 
where   $i, j \in \LAT$
and $\{A, B\}=AB+BA$ (anticommutator).
For each   subset $\I$ of $\LAT$, 
$\AlI$ denotes the subsystem on $\I$ 
 given as the $\cstar$-algebra generated by all 
$\aicr$ and $\ai$  with $i \in \I$.
For $\I\subset \J$, $\AlI$ is naturally imbedded in 
 $\AlJ$ as its subalgebra.

We have already shown that 
SSA  (\ref{eq:SSA}) holds for
  the CAR systems.
 For the convenience, we
  sketch  its    proof 
  given  in  Ref. \citen{Mmaster}
 and Theorem 10.1 of Ref. \citen{AM}.
First we check the
 commuting square property  for a CAR system
 as follows:
\begin{eqnarray*}
 \begin{CD}
\AlIuJ @>\EI>> \AlI\\
@V{\EJ}VV @VV{\EIaJ}V \\
\AlJ @>>\EIaJ>  \AlIaJ,
\end{CD}
\end{eqnarray*}
 where $E_{}$ denotes the conditional expectation 
 with respect to    the tracial state 
onto the subsystem with a specified index.   
 From this property  
  SSA  follows 
for every  state (without any assumption on the state,  like its 
evenness)  by
  a well-known proof method 
 using    the  monotonicity of relative entropy 
 under the action of  conditional expectations.

We  move to  entropy inequalities
 for which CAR makes difference.
The following  
is  usually referred to as the triangle inequality:
\begin{eqnarray}
\label{eq:TRIANGLES}
\Bigl\vert  S(\vpI) -  S(\vpJ)\Bigr\vert 
\le S(\vpIuJ),
\end{eqnarray}
 where $\I$ and $\J$ are disjoint.
While this    is satisfied 
  for   the tensor product systems 
\cite{ARAKILIEB},
it   is not valid  in general 
for  the CAR systems; there is a counter example   
\cite{ENTA}.

We next  introduce our main target,
\begin{eqnarray}
\label{eq:MONOSSA}
S(\vpI)+S(\vpJ)\leq S(\vpKI)+S(\vpKJ),
\end{eqnarray}
 where  $\I$,  $\J$, and $\K$ 
 are disjoint.
We may call (\ref{eq:MONOSSA})   ``MONO-SSA'',
 because it is equivalent to SSA (\ref{eq:SSA})
   for  the tensor product systems  at least,  
 and it  obviously implies  the  monotonicity of the 
 following function 
\begin{eqnarray*}
\K \mapsto S(\vpKI)+S(\vpKJ),
\end{eqnarray*}
 with respect to the inclusion of the index $\K$.
   Our   question  is whether 
 MONO-SSA    holds  for  the CAR systems,
if not, under what  condition    it  is satisfied.

  The  MONO-SSA  for the tensor product   systems
is shown  by 
what is called  purification 
 implying  the equivalence of MONO-SSA and SSA
 for those systems (see 3.3 of Ref. \citen{MBR}). 
 We note that the purification
   is a sort of
 state extension, and   is not   automatic 
for  the CAR systems.
We  shall review     the  basic concept  of 
 state extension. 

In  the description of 
a quantum composite  system, the  total system 
is given  by a $\cstar$-algebra $\Al$, and
its subsystems are described by $\cstar$-subalgebras
$\Al_{i}$ of  $\Al$ indexed by  $i=1,2,\cdots$.   
Let $\vp$
 be a state of $\Al$. We denote 
its  restrictions to $\Ali$ by $\vpi$. Surely
 $\vpi$ is a state of $\Al_i$. 
Conversely, suppose that a set of 
states $\vpi$ of $\Al_i$, $i=1,2,\cdots,$
are   given. Then a state $\vp$  of $\Al$
 is called an  extension of $\{\vpi\}$  
 if its  restriction  to each $\Al_i$
 coincides with  $\vpi$.

 For    tensor product systems, 
 there always exists a  state extension 
 for any given prepared states $\{\vpi\}$ on   disjoint regions,
at least     their   product state extension  
$\vp=\vp_1\otimes \cdots
 \otimes \vp_i\otimes \cdots$,   and  generically  
other extensions.
On the contrary, it 
 is not always the case   for  CAR systems. 
When  two (or  more than two)    prepared states
 on disjoint regions  are not even,
there may be   no   state extension.
We have shown that if all of them are noneven pure states, 
 then there exists  no state extension
\cite{ENTA} \cite{AMEXT}. 

We  explain 
 the   above-mentioned  purification in terms of state extension.
We are     given a state $\vrho_1$
 of  $\AlI$.  We then prepare 
 some state $\vrho_2$ on   some $\AlJ$
 with $\J \cap \I=\emptyset$  
such that it  has  the same nonzero eigenvalues and their 
 multiplicities 
as $\vrho_1$ for  their density matrices.
 We want to  construct 
 their  pure   state extension 
to   $\AlIuJ$.
We    use  the term  
``symmetric purification'' 
 to refer to this procedure where
 the ``symmetric''  may indicate  
the above specified property of  $\vrho_2$.
For  the tensor product systems,
 symmetric purification 
 exists  for every  $\vrhof$.
On the contrary 
 for the CAR systems,
though we can easily  make a pure state extension 
 of $\vrhof$,
   its  pair  $\vrhos$ cannot be always  chosen among  those 
  states which have  the same nonzero eigenvalues and their 
 multiplicities 
as $\vrho_1$.
In the above and what follows, we shall identify  states with their
 density matrices when there is no fear of   confusion.

We  will show  that MONO-SSA  
is not satisfied  in general  in    $\S$ \ref{sec:VIO}.
However  it is shown to hold   for every even state
in $\S$ \ref{sec:SYM}.
{\bf{TABLE 1}} shows  the  truth 
($\bigcirc$) and   the
 falsity ($\times$) of the    
entropy inequalities.\\
\ \\
{\bf{TABLE 1.}} Truth and the falsity of 
von Neumann entropy inequalities\\
\begin{tabular}{|l|c|c|c} \hline 
{Property}& {Tensor-product systems} &{CAR systems} \\ \hline
{SSA}&  $\bigcirc$ & $\bigcirc$ \\ \hline
{Triangle}& $\bigcirc$  &$\times$ in general, but  $\bigcirc$ 
for every even state \\ \hline
{MONO-SSA}& $\bigcirc$  &$\times$ in general, 
but $\bigcirc$ for every even state \\ \hline
\end{tabular}
\ \\

 We  fix  our  notation.
The even-odd  grading 
$\Theta$ is 
determined by
\begin{eqnarray}
\label{eq:THETA}
\Theta(\aicr)=-\aicr, \quad 
\Theta(\ai)=-\ai.  
\end{eqnarray}
The even and odd parts of  $\AlI$ are given  by
\begin{eqnarray*}
\AlIpm 
&\equiv& \Bigl\{ A \in \AlI \, \bigl|\, \Theta(A)=\pm A  \Bigr\}.
\end{eqnarray*}
For an element  $A \in \AlI$ we have the  decomposition
\begin{eqnarray*}
A=\Ap+\Am,\quad 
\Apm\equiv \frac{1}{2}\bigl(A\pm \Theta(A)\bigr)
\in \AlIpm.
\end{eqnarray*} 
For a finite subset $\I$, define 
\begin{eqnarray}
\label{eq:vIEQ}
 \vI \equiv \prod_{i \in \I}\vi,\quad \vi \equiv \aicr\ai-\ai\aicr. 
\end{eqnarray}
By a simple computation,
$\vI$ is   
  a self-adjoint unitary operator in
  $\AlIp$ 
implementing $\Theta$, namely
\begin{eqnarray}
\label{eq:advIin}
   {\mbox {Ad}}(\vI)(A) =\Theta(A),\quad
 A\in \AlI.
\end{eqnarray}
For a finite subset  $\I$,
every even pure state of $\AlI$ 
 is given 
by an eigenvector  of $\vI$
as its  vector state.
 
The following  is  a simple consequence of the CAR 
 given  e.g. in $\S$ 4.5 of Ref. \citen{AM}. 
\begin{lem}
\label{lem:ICOMM}
Let   $\I$ be a finite subset  and   
 $\J$ 
be a (finite or 
infinite) 
subset disjoint with  $\J$.
%
Let 
$\AlIpIJ \equiv \AlIcommut\cap\AlIuJ$
 and 
$\AlIpJ \equiv\AlIcommut\cap\AlJ$,
 the commutant of $\AlI$ in $\AlIuJ$
 and that in $\AlJ$, respectively.
Then
\begin{eqnarray}
\label{eq:IcommIJ}
\AlIpIJ &=&\AlJp + \vI \,\AlJm, \\
\label{eq:IcommJ}
\AlIpJ &=&\AlJp.
\end{eqnarray}
\end{lem}

In this note we   restrict our discussion  to
 finite-dimensional  systems 
so as to exclude 
 from the outset 
 the  cases where
our statements themselves 
on  von Neumann entropy 
 do not make sense;
  for   infinite-dimensional systems  
  a density matrix  does not exist in general for a given state.
(However in the proof of Proposition \ref{pro:vrhostil}
we shall mention  
   possible infinite-dimensional 
 extensions  of some  results.)

\section{Symmetric Purification for Even States}
\label{sec:SYM}
Symmetric purification
 is   a  useful    mathematical  
 technique having a  lot  of applications.
For example,  we can  derive   
MONO-SSA from SSA  
for the tensor product systems  by using  it.

We now discuss  symmetric purification
 for the CAR systems.
We shall show its  existence  for even states.
\begin{lem}
\label{lem:entropysymm}
Let $\I$ and $\J$ be  mutually disjoint finite subsets. 
Let $\vrho$  be an even  pure state 
 of $\AlIuJ$, and let
 $\vrhof$ and  $\vrhos$ be its  restrictions
 to  $\AlI$ and $\AlJ$.
Then  the density matrix of $\vrhof$
 has the  same nonzero eigenvalues and their 
 multiplicities 
as those  of $\vrhos$.
   In particular, 
$S(\vrhof)=
S(\vrhos)$.
\end{lem}
\proof
 We have 
$\AlIuJ=\AlI\otimes \AlIpIJ$ by (\ref{eq:IcommIJ}). 
 By  some  finite-dimensional Hilbert spaces
  $\Hilf$, $\Hils$ and  $\Hilfs\equiv \Hilf\otimes \Hils$,
we can write  $\AlIuJ=\Bl(\Hilfs)$,
 $\AlI= \Bl(\Hilf)$, and  $\AlIpIJ=\Bl(\Hils)$.

Since $\vrho$ is a pure state of $\AlIuJ$,
 its density matrix 
 (with respect to the non-normalized trace $\Tr$ of $\Bl(\Hilfs)$)
is a one-dimensional projection operator 
 of $\Bl(\Hilfs)$, and hence there exists
 a unit vector
$\xi\in \Hilfs$     
such that $D_{\vrho}\eta=(\xi, \eta)\xi$
 for any $\eta\in \Hilfs$. 
By using the Schmidt decomposition \cite{SCH}, we have 
 the following decomposed form:
\begin{eqnarray}
\label{eq:xidef}
\xi=\sum_{i} \lambda_{i} \xi_{1i}\otimes \xi_{2i},\quad \lambda_i>0,
\end{eqnarray}
where $\{\xi_{1i}\}$ and  $\{\xi_{2i}\}$
 are some orthonormal sets of vectors of $\Hilf$
 and $\Hils$.
For $\nu=1,\,2$,
let $P(\xi_{\nu i})$ denote  the projection operator 
 on the one-dimensional subspace of $\Hil_{\nu}$
 containing $\xi_{\nu i}$.
We  denote the restricted states of $\vrho$
 onto  $\Bl(\Hils)$
 by   $\vrhostil$.
By (\ref{eq:xidef}), the density matrices of $\vrhof$ and $\vrhostil$
 have the following symmetric forms: 
\begin{eqnarray}
\label{eq:dv1}
D_{\vrhof}=\sum_{i} \lam_{i}^{2}P(\xi_{1 i} ),\quad 
 D_{{\vrhostil}}=\sum_{i} \lam_{i}^{2}P(\xi_{2 i} ).
\end{eqnarray}

Since $\vrho$ is an even state,
 its restriction  $\vrhos$
  is   even
 and hence its density matrix 
  $D_{\vrhos}$
belongs to 
 $\AlJp$.

On the other hand, the even state $\vrho$
 is invariant under
 the action of ${\mbox {Ad}}(\vIuJ)={\mbox {Ad}}(\vI\vJ)$:
\begin{eqnarray*}
\label{eq:}
\vIuJ D_{\vrho} \vIuJ=\vI\vJ D_{\vrho}\vI\vJ
= D_{\vrho}.
\end{eqnarray*}
Acting  the conditional expectation
 onto $\Bl(\Hils)$ with respect to the tracial state
 of $\Bl(\Hilfs)$ on the above equality,
 we obtain
\begin{eqnarray*}
\label{eq:}
\vJ D_{\vrhostil}\vJ= D_{\vrhostil}
\end{eqnarray*}
noting that  $\vI$ belongs to  $\Bl(\Hilf)=\Bl(\Hils)^{\prime}
\cap\Bl(\Hilfs)$.   

We denote  $\Bl(\Hils)_{+}\equiv\Bl(\Hils)\cap \AlIuJp$, 
  the set of all invariant elements under 
${\mbox {Ad}}(\vI\vJ)$ in $\Bl(\Hils)$.
By (\ref{eq:IcommIJ}), $\Bl(\Hils)_{+}$ is equal to $\AlJp$, and
 also to 
the set of all invariant elements under 
${\mbox {Ad}}(\vJ)$ in $\Bl(\Hils)$.
Therefore  both
$D_{\vrhostil}$ and   $D_{\vrhos}$ belong to
 $\Bl(\Hils)_{+}$.
Accordingly, $D_{\vrhostil}$ is equal to  $D_{\vrhos}$
 as the density of the state $\vrho$ 
 restricted to $\Bl(\Hils)_{+}$, and hence 
\begin{eqnarray}
\label{eq:dv2}
 D_{\vrhos}=
D_{\vrhostil}=\sum_{i} \lam_{i}^{2}P(\xi_{2 i} ).
\end{eqnarray}

From (\ref{eq:dv1}) and (\ref{eq:dv2}),  
it follows that 
 $\vrhof$ and  $\vrhos$
have the  same nonzero eigenvalues and their 
 multiplicities  
 equal to $\{\lam_{i}^{2} \}$.
Thus
\begin{eqnarray*}
S(\vrhof)=
S(\vrhos).
\end{eqnarray*}
\proofend\\

For a subset $\I$ of $\LAT$,
$|\I|$ denotes  the number of sites in  $\I$.
\begin{lem}
\label{lem:Iext}
Let $\I$ be a finite subset
and $\vrhof$ be a  state of $\AlI$.
Let $\J$ be a finite subset 
such that $\J\cap\I=\emptyset$ and $|\J|\ge|\I|$. 
Then there exists  a pure state  $\vrho$  on   $\AlIuJ$
 satisfying
\begin{eqnarray}
\label{eq:}
\vrho|_{\AlI}=\vrhof.
\end{eqnarray}
Moreover, if $\vrhof$ is even,
 then the above $\vrho$ can be taken to be  even.   
\end{lem}
\proof 
We  use the same notation as  in the   
 proof of the  preceding lemma and 
write   $\AlIuJ=\Bl(\Hilfs)$,
 $\AlI= \Bl(\Hilf)$, and  $\AlIpIJ=\Bl(\Hils)$.
Let $\vrhof=\sum_{i} \lam_{i}^{2} P(\xi_{1 i} )$,
 where $\lam_i>0$, $\{\xi_{1i}\}$ is 
 an orthonormal set of  $\Hilf$, and 
 $P(\xi_{1 i} )$ is  the  projection operator
 on the one-dimensional subspace of $\Hilf$
 containing $\xi_{1 i}$.
Since $|\J|\ge|\I|$ and hence 
${\mbox{dim}}\; \Hils\ge {\mbox{dim}}\; \Hilf$,  
we can take
  an   orthonormal set
 of vectors  $\{\xi_{2i}\}$
of $\Hils$ having  the same cardinality as 
$\{\xi_{1i}\}$.
Define a unit vector 
$\xi \in \Hilfs$ by the same formula 
as (\ref{eq:xidef}) and let 
$\vrho$ be its   vector  state,
 namely the state whose density matrix is 
the  projection operator
 on the one-dimensional subspace of $\Hilfs$
 containing $\xi$.
This $\vrho$ is a pure state extension  
 of $\vrhof$ to $\AlIuJ$ 
 by its 
definition.

Assume now 
 that $\vrhof$ is even, and
hence  its density matrix is in 
$\AlIp$. For each  eigenvalue, the associated 
spectral projection 
 is also even and  commutes with $\vI$, 
 and its  range is invariant under $\vI$.
Therefore
 we can choose an orthonormal basis
 of the range of the  projection 
which consists of eigenvectors of $\vI$.
  We take   $\{\xi_{1i}\}$
 to be a set of  eigenvectors  of $\vI$.

Since 
 $\vJ$ belongs to $\Bl(\Hils)_{+}(=\AlJp)$, there 
 exists an orthonormal basis of $\Hils$ consisting of 
 eigenvectors of $\vJ$.  
Due to the assumption   $|\J|\ge|\I|$,
we can take  
  a set of different eigenvectors 
$\{\xi_{2i}\}$ of  $\vJ$ such that for each $i$ its  
 eigenvalue, $+1$ or $-1$, is equal to that
  of  $\xi_{1i}$ for $\vI$. 
Define  a unit  vector $\xi$ by   (\ref{eq:xidef})
  using these 
$\{\xi_{1i}\}$ and $\{\xi_{2i}\}$.
 Since this $\xi$ is an eigenvector of $\vIuJ$ by its definition,
 its  vector  state $\vrho$ is  even. 
\proofend\\

Combining the above two lemmas
 we obtain the following.
\begin{pro}
\label{pro:symmext}
Let $\I$ be a finite subset and 
 $\vrhof$ be an even  state of $\AlI$.
Let $\J$ be a finite subset 
such that $\J\cap\I=\emptyset$ and $|\J|\ge|\I|$. 
 Then there exists  
  an even pure  state $\vrho$ 
 on  $\AlIuJ$ such that its restriction to 
 $\AlI$
  is equal to  $\vrhof$ and 
the density matrix of its restricted state 
$\vrhos\equiv\vrho|_{\AlJ}$
 has the  same nonzero eigenvalues and their 
 multiplicities 
as those  of $\vrhof$.
\end{pro} 

We may call  the above state extension
  from $\vrhof$ to $\vrho$ 
the  symmetric purification. 
Thanks to  this,
 we obtain the following two theorems.
 \begin{thm}
\label{thm:MONO-SSAeven}
Let   $\I$,  $\J$ and $\K$ 
 be mutually disjoint  finite subsets.
For every even state $\vp$, MONO-SSA 
\begin{eqnarray}
\label{eq:thmMONOSSA}
S(\vpI)+S(\vpJ)\leq S(\vpKI)+S(\vpKJ)
\end{eqnarray}
is satisfied.
\end{thm}
\proof 
The equivalence of MONO-SSA and SSA 
 for  even states follows
 from Proposition \ref{pro:symmext}  
  in the same way as 
(3) p164 of Ref. \citen{ARAKILIEB}.
Since SSA holds 
for every state, 
MONO-SSA is  valid for every even state.
\proofend\\
\ \\
Similarly,  by using  Proposition \ref{pro:symmext}
 we  immediately obtain   the 
triangle inequality for even states
 in much the same way 
as (3.1) of Ref. \citen{ARAKILIEB}.
We omit its  proof.
\begin{thm}
\label{thm:triangle}
Let  $\I$ and $\J$ be mutually 
 disjoint finite subsets.
For every even state $\vp$, the triangle inequality 
\begin{eqnarray}
\label{eq:thmTRIANGLES}
\Bigl\vert  S(\vpI) -  S(\vpJ)\Bigr\vert 
\le S(\vpIuJ)
\end{eqnarray}
 holds.
\end{thm}
%
%
\section{Violation of MONO-SSA}
\label{sec:VIO}
In this section we give    a certain class of 
noneven states.
\ \\
\hspace{1cm}
\unitlength 0.1in
\begin{picture}( 39.2000,  6.9000)(  2.1000, -9.3000)
%
\special{pn 13}%
\special{pa 410 610}%
\special{pa 1410 610}%
\special{fp}%
\special{sh 1}%
\special{pa 1410 610}%
\special{pa 1344 590}%
\special{pa 1358 610}%
\special{pa 1344 630}%
\special{pa 1410 610}%
\special{fp}%
%
\special{pn 13}%
\special{pa 1390 610}%
\special{pa 420 610}%
\special{fp}%
\special{sh 1}%
\special{pa 420 610}%
\special{pa 488 630}%
\special{pa 474 610}%
\special{pa 488 590}%
\special{pa 420 610}%
\special{fp}%
%
\special{pn 13}%
\special{pa 1580 610}%
\special{pa 2590 610}%
\special{fp}%
\special{sh 1}%
\special{pa 2590 610}%
\special{pa 2524 590}%
\special{pa 2538 610}%
\special{pa 2524 630}%
\special{pa 2590 610}%
\special{fp}%
%
\special{pn 13}%
\special{pa 2490 610}%
\special{pa 1480 600}%
\special{fp}%
\special{sh 1}%
\special{pa 1480 600}%
\special{pa 1546 622}%
\special{pa 1534 602}%
\special{pa 1548 582}%
\special{pa 1480 600}%
\special{fp}%
%
\special{pn 13}%
\special{ar 1440 574 1230 290  4.9120284 6.2831853}%
\special{ar 1440 574 1230 290  0.0000000 3.9619712}%
\put(6.9000,-4.1000){\makebox(0,0)[lb]{pure on $\I\cup\K$}}%
\put(6.3000,-5.7000){\makebox(0,0)[lb]{tracial}}%
\put(16.6000,-5.9000){\makebox(0,0)[lb]{pure, noneven}}%
\put(7.2000,-7.4000){\makebox(0,0)[lb]{$\I$}}%
\put(18.6000,-7.3000){\makebox(0,0)[lb]{$\K$}}%
\put(27.9000,-6.0000){\makebox(0,0)[lb]{$\circ$}}%
\put(34.4000,-5.6000){\makebox(0,0)[lb]{even}}%
\put(34.4000,-7.9000){\makebox(0,0)[lb]{$\J$}}%
\put(3.7000,-11.0000){\makebox(0,0)[lb]{FIG.1. A state not satisfying
 MONO-SSA. }}%
%
\special{pn 13}%
\special{pa 2950 610}%
\special{pa 2890 600}%
\special{fp}%
\special{sh 1}%
\special{pa 2890 600}%
\special{pa 2952 632}%
\special{pa 2944 610}%
\special{pa 2960 592}%
\special{pa 2890 600}%
\special{fp}%
%
\special{pn 13}%
\special{pa 2970 610}%
\special{pa 4130 600}%
\special{fp}%
\special{sh 1}%
\special{pa 4130 600}%
\special{pa 4064 582}%
\special{pa 4078 600}%
\special{pa 4064 622}%
\special{pa 4130 600}%
\special{fp}%
\end{picture}%
\ \\
\ \\
\ \\
 
We  shall give a  sketch of our model  indicated by  FIG.1.
We can take a pure state $\vrho_{\IuK}$ on $\I\cup\K$ 
 whose restriction $\vrho_{\K}$ 
is a pure state,
 but $\vrho_{\I}$ is non-pure,  say the tracial state. 
 Such  $\vrho_{\IuK}$ does not satisfy 
the triangle inequality, 
 because the entropies on $\I$ and on $\K$ 
 are different,  whereas the entropy  on $\I\cup\K$
 is zero.
 It can be said that  
 the pure state $\vrho_{\IuK}$
 has  the asymmetric restrictions 
  in our  terminology.
This asymmetry is due to the  
large  amount of   the 
 oddness of  $\vrho_\K$, whose  precise  meaning  
will be given soon.
(Note however that  for the infinite-dimensional case,
 the GNS representations 
$\pivrhoK$ and $\pivrhoKt$ 
should be  unitarily equivalent, 
see Proposition \ref{pro:vrhostil}
 (i).)
We take an arbitrary even state $\vrho_{\J}$
 on $\J$.
The desired state on $\vrho_{\I\cup\K\cup\J}$
 on $\I\cup\K\cup\J$ is given by the 
product state extension of 
$\vrho_{\I\cup\K}$
and $\vrho_{\J}$, which will be denoted by 
$\vrho_{\I\cup\K}\circ\vrhoJ$.

We   recall the definition of the transition 
probability  \cite{UHLMANN77}.
For two states $\vp$ and $\psi$ 
 of  $\AlI$ (where 
$|\I|$ is  finite or infinite),
 take  any representation $\pi$ of $\AlI$ on a Hilbert space
 $\Hil$ containing  vectors $\phivec$ and  $\psivec$
 such that 
\begin{eqnarray}
\label{eq:phivecEQ}
\vp(A)=(\phivec,\, \pi(A) \phivec),\quad
\psi(A)=(\psivec,\, \pi(A)\psivec),
\end{eqnarray}
for all $A\in \AlI$.
The transition probability between 
 $\vp$ and $\psi$ is given    by
\begin{eqnarray}
\label{eq:Trans0}
P(\vp,\,\psi)\equiv
\sup | ({\phivec},\,\psivec) |^{2},
\end{eqnarray}
where the supremum is taken over all $\Hil$, $\pi$, 
 $\phivec$ and $\psivec$ as described 
 above.
 For a state $\vp$ of $\AlI$, we define 
\begin{eqnarray}
\label{eq:pT}
p_{\Theta}(\vp)\equiv P(\vp,\,\vp\Theta)^{1/2},
\end{eqnarray}
where 
$\vp\Theta$ denotes the state 
$\vp \Theta(A)=\vp(\Theta(A))$, $A\in \AlI$.
Intuitively, $p_{\Theta}(\vp)$ quantifies the amount of {\it{oddness}} 
 of the state $\vp$. If $p_{\Theta}(\vp)=0$ or nearby,
 then we  may say that 
 the difference between 
$\vp$ and $\vp\Theta$  is large.
If $\vp$ is even, $p_{\Theta}(\vp)$ takes obviously the maximum 
 value 1.

The following  
 is  Lemma 3.1 of Ref. \citen{AM}. 
\begin{lem}
\label{lem:uone}
If $\vrhof$ is a pure state of $\AlK$
and  $\pivrhof$ and $\pivrhoft$ are unitarily 
 equivalent, then there exists a self-adjoint  unitary 
 $\uone \in \pi_{\vrhof}(\AlKp)^{\prime \prime}$ satisfying 
\begin{eqnarray} 
\label{eq:uonerel}
\uone \pivrhof(A) \uone=\pivrhof  (\Theta(A)),\quad A\in \AlK.
\end{eqnarray}
\end{lem} 

The next proposition 
 is a  basis of our construction.
  It  is a generalization 
 of   Ref. \citen{ENTA}.  The  
first paragraph  is in  principle
  excerpted     from 
 Theorem 4 (4) and (5) 
of Ref. \citen{AM}. 
The second paragraph 
  is necessary for  the argument of  entropy.
\begin{pro}
\label{pro:vrhostil}
Let $\K$ and $\I$ be  mutually disjoint  subsets. 
Assume that  $\vrhof$  is a (noneven) pure state 
of $\AlK$ satisfying  $p_{\Theta}(\vrhof)=0$. Assume that  
 $\vrhos$ is an even state of $\AlI$. There exists a joint extension 
 of $\vrhof$ and $\vrhos$ other than  
their product state extension
 if and only if $\vrhof$ and $\vrhos$ satisfy 
 the following pair of conditions$:$\\
\quad {\rm{(i)}} $\pivrhof$ and $\pivrhoft$ 
are unitarily equivalent.\\
\quad {\rm{(ii)}} There exists a state $\vrhostil$ 
 of $\AlI$ such that $\vrhostil \ne \vrhostil\Theta$
 and 
\begin{eqnarray}
\label{eq:2dec}
\vrhos=\frac{1}{2}\bigl(\vrhostil +\vrhostil\Theta  \bigr).
\end{eqnarray}
For  each $\vrhostil$
 above, there exists the  joint extension  
 of $\vrhof$ and $\vrhos$
to $\AlKuI$ denoted by 
$\psi_{\vrhostil}$
which satisfies
\begin{eqnarray}
\label{eq:psivrhostil}
\psi_{\vrhostil}(\Af \As)=\vrhof(\Af)\vrhos(\Asp)+
\vrhofo(\pivrhof(\Af)\uone)\vrhostil(\Asm),  
\end{eqnarray}
 where $\vrhofo$
 is the GNS-extension of 
  $\vrhof$ to $\pi_{\vrhof}(\AlK)^{\prime \prime}$.

If $\K$ and  $\I$ are   finite subsets,
then the entropy  of 
 ${\vrhostil}$ is equal to that  of $\psi_{\vrhostil}$.
\end{pro}
\proof
We shall  show   only  the sufficiency  of 
the pair of conditions  (i) and (ii).
For the necessity of (i)  see 5.2 in Ref. \citen{AM},
 and for that of (ii)  see (d)
 in the proof of its Theorem 4(4).

Let $(\Hilvrhof$, $\pivrhof$, $\Vecvrhof)$ 
 be  a GNS triplet for $\vrhof$  and
 $(\Hilvrhostil$, $\pivrhostil$, $\Vecvrhostil)$
 be that for $\vrhostil$.
Define 
\begin{eqnarray}
\label{eq:piform1}
\Hil&\equiv& \Hilvrhof \otimes \Hilvrhostil,\quad 
\Vec\equiv \Vecvrhof\otimes\Vecvrhostil, \\
\label{eq:piform2}
\pi(\Af \As)&\equiv& \pivrhof(\Af)\otimes 
\pivrhostil(\Asp)+
 \pivrhof(\Af)\uone \otimes \pivrhostil(\Asm),
\end{eqnarray}
for $\Af\in \AlK$, $\As=\Asp+\Asm$, 
$\Aspm \in \AlIpm$.
Let $\id_1$ be the identity operator of 
$\Hilvrhof$ and $\id_2$ be that of $\Hilvrhostil$.
 
We can check  that the operators $\pi( \AlKuI)$ satisfy the CAR 
  by  using (\ref{eq:uonerel}), and hence
 $\pi$ extends to a representation of 
  $\AlKuI$.
  We define the state $\psi_{\vrhostil}$
on $\AlKuI$ as
\begin{eqnarray}
\psi_{\vrhostil}(A)\equiv    
(\Vec,\,\pi(A)\Vec)
\end{eqnarray}
 for $A\in \AlKuI$.
The von Neumann algebra $\pi( \AlKuI)^{\prime\prime}$
 is generated by 
$\pivrhof(\AlK)^{\prime\prime}\otimes\id_2$, 
 $\id_{1} \otimes \pivrhostil(\AlIp)^{\prime\prime}$,
 and  the weak closure of $\id_{1} \otimes \pivrhostil(\AlIm)$,
 where 
we have noted 
$\uone \in \pi_{\vrhof}(\AlKp)^{\prime \prime}=\Bl(\Hilvrhof)$.
Therefore  
\begin{eqnarray}
\label{eq:tensorkata}
\pi( \AlKuI)^{\prime\prime}=
\Bl(\Hilvrhof)\otimes  \pivrhostil(\AlI)^{\prime\prime}.
\end{eqnarray}
From this it follows that 
the vector $\Vec$ is cyclic for the representation  $\pi$ 
of  $\AlKuI$ in $\Hil$.
Hence  $(\Hil$, $\pi$, $\Vec)$ 
 gives   a GNS triplet for the state 
$\psi_{\vrhostil}$ on $\AlKuI$. 

 We have
 \begin{eqnarray}
\label{eq:bunkai}
\psi_{\vrhostil}(\Af\As)=
(\Vec,\,\pi(\Af\As)\Vec)=
\vrhof(\Af)\vrhostil(\Asp)+
 \vrhofb(\pivrhof(\Af)\uone)\vrhostil(\Asm),
\end{eqnarray}
 for $\Af\in \AlK$, $\As=\Asp+\Asm$, 
$\Aspm \in \AlIpm$.
Taking $\As=\id$ in (\ref{eq:bunkai}),
we  obtain
\begin{eqnarray}
\label{eq:fitt}
\psi_{\vrhostil}(\Af)=\vrhof(\Af).    
\end{eqnarray}

We will then show 
\begin{eqnarray}
\label{eq:sitt}
\psi_{\vrhostil}(\As)=\vrhos(\As).    
\end{eqnarray}
Under the condition of Lemma \ref{lem:uone} (which is our case),
 we have 
\begin{eqnarray}
\label{eq:ptone}
p_{\Theta}(\vrhof)=
|\vrhofb(\uone)|,
\end{eqnarray}
because the transition probability
 between  the  vector states
 of  the algebra $\BHvrhof=\pivrhof(\AlK)^{\prime\prime}$ is 
 equal to the (usual) 
transition probability of their vectors and
 hence 
$p_{\Theta}(\vrhof)=
    | (\Vecvrhof,\,\uone \Vecvrhof) |$.
 By the assumption $p_{\Theta}(\vrhof)=0$, (\ref{eq:ptone})
 implies
 \begin{eqnarray}
\label{eq:kieru}
\vrhofb(\uone)=0.
\end{eqnarray}
Setting $\Af=\id$ in (\ref{eq:bunkai}),
 we obtain 
 \begin{eqnarray}
\label{eq:exthe1}
\psi_{\vrhostil}(\As)&=&
\vrhostil(\Asp)+
 \vrhofb(\uone)\vrhostil(\Asm)\nonum \\
&=&\vrhostil(\Asp).
\end{eqnarray}
By (\ref{eq:2dec}),
\begin{eqnarray}
\label{eq:exthe2}
\vrhos(\As)
=\vrhos(\Asp)
=\vrhostil(\Asp).
\end{eqnarray}
From (\ref{eq:exthe1}) and (\ref{eq:exthe2}), 
(\ref{eq:sitt}) follows.
We have  now shown that $\psi_{\vrhostil}$
 is an extension of $\vrhof$ and $\vrhos$.  

We will show the  second paragraph.
By (\ref{eq:tensorkata})
 and  the commutant theorem,
$\pi( \AlKuI)^{\prime}=
\id_{1}\otimes\pivrhostil(\AlI)^{\prime}$,
  where the commutant is taken in
 each GNS space.
Thus we have the  following isomorphism: 
\begin{eqnarray}
\label{eq:isomor}
 b \mapsto
 \id_1\otimes b,\quad b\in \pivrhostil(\AlI)^{\prime}
  \end{eqnarray}
 from $\pivrhostil(\AlI)^{\prime}$ onto
$\pi(\AlKuI)^{\prime}$.
Furthermore
by (\ref{eq:piform1})  we obtain
\begin{eqnarray}
\label{eq:weight}
(\Vec,\,  (\id_1\otimes b)\Vec)=
(\Vecvrhof\otimes\Vecvrhostil,\,  
\Vecvrhof\otimes b \Vecvrhostil)
=(\Vecvrhostil,\,  b\Vecvrhostil).
\end{eqnarray}

From  
 the assumption 
 that $\K$ and $\I$ are finite subsets (which 
 we have not used so far), 
 $\pivrhostil(\AlI)^{\prime}$ and 
$\pi(\AlKuI)^{\prime}$ are both
 finite-dimensional type I factors.

Now we  note  the following  basic fact about 
 GNS representations for  states  of  finite-dimensional 
 type I factors which  can be considered as 
 the  counterpart
 of Lemma \ref{lem:entropysymm} for the usual case, namely 
for a pair of isomorphic systems coupled by tensor product.
Let $\ome$ be a state of a 
 finite-dimensional 
 type I factor 
$\Almath$ and   $(\Hilome$, $\piome$, $\Vecome)$ 
 denote    a GNS triplet of $\ome$.
The GNS vector  $\Vecome$ of $\ome$
 induces a state on the commutant 
$\piome(\Almath)^{\prime}$ 
whose  expectation value 
for $a\in 
\piome(\Almath)^{\prime}$ 
 is given by 
$(\Vecome,\,  a\Vecome)$. We call this state 
on $\piome(\Almath)^{\prime}$ 
``$\ome$ on the commutant''.
(In our terminology,  the pure  state 
 with respect to  $\Vecome$ on $\Bl(\Hilome)=
\piome(\Almath)\otimes 
\piome(\Almath)^{\prime}$
gives a symmetric purification of $\ome$.  
Also  $\ome$ on the commutant is symmetric to $\ome$.)
Then the entropy of $\ome$ on $\Almath$ (equivalently 
 that on  $\piome(\Almath)$)
 is equal to the entropy  of $\ome$ on the commutant
 by the same reason described in Lemma \ref{lem:entropysymm}. 
We  note  that this 
holds for a general $\cstar$-algebra
 if the  GNS representation of a given state generates  
a type I von Neumann algebra with a discrete center.
Similarly the extension of 
 Lemma \ref{lem:entropysymm} 
 is possible  under the above condition on the state. 

From  the above  fact  with  
(\ref{eq:isomor}) and 
(\ref{eq:weight}),
we  deduce  the equality of  the entropies 
 of  ${\vrhostil}$  and  of $\psi_{\vrhostil}$
\proofend\\
\ \\
{\it{{Remark 1}}}:\
Note that this  $\psi_{\vrhostil}$
 is a state extension of $\vrhof$ and $\vrhos$,
 not that of $\vrhof$ and $\vrhostil$.
The possibility of the state extension of 
 $\vrhof$ and $\vrhostil$ is negated by Theorem 4 (3)
 of Ref. \citen{AMEXT}.\\
\ \\
{\it{{Remark 2}}}:\
We  can easily   make 
 examples of states in this  proposition.
Take a finite subset $\K$
 and  an odd self-adjoint element  $A$
  in the algebra $\AlK$ which we will  identify  with  
$\Bl(\Hil)$ on a finite-dimensional Hilbert space $\Hil$.
Let $\eta\in \Hil$ be a normalized  eigenvector
 of this $A$ and $\ome_\eta$ denote  the associated   vector state.
Then  
$\eta\perp \vK\eta$, and 
$\ome_\eta\Theta$ becomes 
the 
 vector state with respect to 
$\vK\eta$. Hence
 $p_{\Theta}(\ome_\eta)=0$. This
$\ome_\eta$ obviously   satisfies 
 (i). 
For the  existence of 
 $\vrhos$  satisfying (ii),
  take for example 
 the above $\ome_\eta$ for $\vrhostil$ (with  $i\in \I$),
 $\ome_\eta\Theta$
 for $\vrhostil\Theta$, and
  their  affine sum  (\ref{eq:2dec})
 for $\vrhos$. \ \\

Proposition \ref{pro:vrhostil}
 yields  the following construction    
giving   counter examples of    MONO-SSA
 and those  of the triangle inequality.
\begin{thm}
\label{thm:MONOSSA-VIO}
Let $\K$, $\I$,  and $\J$ be  mutually 
 disjoint  subsets. 
Let $\vrhoK=\vrhof$ and $\vrhoI=\vrhos$ 
 where  $\vrhof$ and $\vrhos$ are those  
 states on $\AlK$
 and on $\AlI$ given in   
Proposition \ref{pro:vrhostil}.
 Let $\vrhoKI$
 be the state extension 
of     $\vrhoK$ and $\vrhoI$ to  $\AlKuJ$
 given by  $\psi_{\vrhostil}$ in the form of 
(\ref{eq:psivrhostil}).
Let $\vrhoJ$ be an arbitrary even state
 of $\AlJ$. 
Then 
 for   such $\vrhoKI$ and $\vrhoJ$,
there exists
 a  (unique) product state extension $\vrhoKI\circ\vrhoJ$ 
on $\AlKIJ$.
If all  $\K$, $\I$, and $\J$ are  finite subsets,
 then
\begin{eqnarray}
\label{eq:MONOVIO}
S(\vrho_{\KuI})+S(\vrho_{\KuJ})
< S(\vrho_{\I})+S(\vrho_{\J}),
\end{eqnarray}
and 
\begin{eqnarray}
\label{eq:TRIVIO}
\Bigl\vert  S(\vrho_\I) -  S(\vrho_\K)\Bigr\vert
=  S(\vrhoI)
> S(\vrho_{\KuI}).
\end{eqnarray}
\end{thm}
\proof
By Ref. \citen{POWERS} or Theorem 1 (1) of Ref. \citen{AM},
  there exists a unique product state extension
 of a pair of states on  disjoint regions
 if (and only if) at least one of them is even, and hence
the (unique) existence of  
$\vrhoKI\circ\vrhoJ$  follows.

By the second  paragraph  
 of Proposition \ref{pro:vrhostil},
we obtain 
\begin{eqnarray}
\label{eq:MONOVIO1}
S(\vrho_{\KuI})=S(\psi_{\vrhostil})
=S(\vrhostil).
\end{eqnarray}
Since  $\vrhos=\frac{1}{2}\bigl
(\vrhostil +\vrhostil\Theta  \bigr)$ and 
$ \vrhostil \ne \vrhostil\Theta$, 
 we have
\begin{eqnarray}
\label{eq:MONOVIO3}
S(\vrho_{\I})=S(\vrhos)=
S\bigl(1/2(\vrhostil +\vrhostil\Theta  ) \bigr)
> 1/2 S(\vrhostil) +1/2S(\vrhostil\Theta )
\end{eqnarray}
by  the strict concavity of von Neumann
 entropy with respect to the affine sum of states (see  
   {\it{{Remark 3}}} below). 
 Due to the unitary invariance of  von Neumann
 entropy, 
\begin{eqnarray}
\label{eq:MONOVIO4}
 S(\vrhostil) =S(\vrhostil\Theta ).
\end{eqnarray}
By (\ref{eq:MONOVIO1}), (\ref{eq:MONOVIO3})
 and (\ref{eq:MONOVIO4}), we have
\begin{eqnarray}
\label{eq:MONOVIO5}
S(\vrhoI)
>  S(\vrhoKI).
\end{eqnarray}
By  the product property of $\vrhoKJ=\vrhoK\circ \vrhoJ$
 the density matrix of $\vrhoKJ$ is a product 
 of the density matrices    of $\vrhoK$  and  $\vrhoJ$
 which mutually commute because $\vrhoJ\in \AlJp$.
 Hence by  a direct computation we  have
 \begin{eqnarray*}
\label{eq:MONOVIO6}
S(\vrhoKJ)=S(\vrhoK)+S(\vrhoJ).
\end{eqnarray*}
Since  $\vrhoK$ is assumed to be pure and hence $S(\vrhoK)=0$,
 we have 
 \begin{eqnarray}
\label{eq:MONOVIO7}
S(\vrhoKJ)=S(\vrhoJ).
\end{eqnarray}
From  (\ref{eq:MONOVIO5}) and (\ref{eq:MONOVIO7}),
 we obtain (\ref{eq:MONOVIO}). 
From   (\ref{eq:MONOVIO5}) and  $S(\vrhoK)=0$, we obtain   
 (\ref{eq:TRIVIO}).
\proofend\\
\ \\
{\it{{Remark 3}}}:\
Let $\Hil$ be a finite-dimensional Hilbert space.  
 For states  $\vp$ and $\psi$
  on 
the algebra  $\Bl(\Hil)$ and for  
$0\le \lambda\le 1$,
the following  von Neumann
 entropy 
inequalities    
 are   well-known:
 \begin{eqnarray}
\label{eq:concavity}
S(\lam \vp+(1-\lam)\psi)\ge \lam S(\vp)+(1-\lam)S(\psi),
\end{eqnarray}
 \begin{eqnarray}
\label{eq:convexity}
S(\lam \vp+(1-\lam)\psi)\le  \lam S(\vp)+(1-\lam)S(\psi)
-\lam\log\lam- (1-\lam)\log(1-\lam).
\end{eqnarray}
We refer to 
  Proposition 6.2.25
 of Ref. \citen{BRA2} for their  proofs.
We now  see  the strict
  concavity of von Neumann entropy which was used 
 in the proof of Theorem \ref{thm:MONOSSA-VIO},
 namely  for  $0<\lam<1$ 
the equality of (\ref{eq:concavity}) holds if and only if 
  $\vp=\psi$. We  employ  the proof method 
 given in the above-mentioned reference.
Let $\Kil$ be a two-dimensional Hilbert space
 and $P$ denote a one-dimensional projection of $\Bl(\Kil)$.
We denote $\Al_{1}\equiv\Bl(\Hil)$,  $\Al_{2}\equiv\Bl(\Kil)$, and 
  $\Al_{1,2}\equiv\Bl(\Hil\otimes \Kil)$.
Let $\ome$ denote  a state on $\Al_{1,2}$
 whose   density matrix $D_{\ome}$ is given by 
 \begin{eqnarray*}
\label{eq:}
\lam D_{\vp}\otimes P+ (1-\lam) D_{\psi}\otimes (1-P). 
\end{eqnarray*}
We denote  $\ome$ restricted to $\Al_{1}$
 (to $\Al_{2}$, respectively)
 by $\omef$ (and $\omes$).
We see  that $\omef$
 is equal to $\lam \vp+(1-\lam)\psi$, hence 
 \begin{eqnarray*}
\label{eq:}
S(\omef)=S (\lam \vp+(1-\lam)\psi).
\end{eqnarray*}
Also we have 
 \begin{eqnarray*}
\label{eq:}
S(\ome)-
S(\omes)=
\lam S (\vp)+(1-\lam)S(\psi).
\end{eqnarray*}
Since   
 \begin{eqnarray*}
\label{eq:}
S(\omef)+
S(\omes)-S(\ome)=
\ome(\log D_{\ome}- \log (D_{\omef}\otimes D_{\omes}) )
\equiv S(\ome| \omef\otimes \omes)
\ge  0,
\end{eqnarray*}
which is equivalent to the    subadditivity of entropy,
we obtain (\ref{eq:concavity}).
 This  $S(\ome| \omef\otimes \omes)$
 is  
 relative entropy 
of the two states in its argument 
and is known to have  
    strict  positivity, 
i.e.  $S(\ome| \omef\otimes \omes)=0$
 if and only if $\ome=\omef\otimes \omes$, 
 which is equivalent to $\vp=\psi$ for  $0<\lam<1$. 
Hence
   our desired    strictness of (\ref{eq:concavity}) is  shown.
 \\
\ \\
{\it{{Remark 4}}}:\
We shall  give a  rough  estimation 
of 
the  amount of  violation of the triangle 
inequality,
 i.e. 
$\vert  S(\vp_\I) -  S(\vp_\K)\vert-S(\vp_{\KuI})$
 of (\ref{eq:TRIVIO})
  for a general state  $\vp$. 
Let 
${\hat{\vp}}\equiv 1/2(\vp+ \vp \Theta)$.
Then  
it obviously satisfies the triangle inequality.
By   (\ref{eq:concavity}) and (\ref{eq:convexity}),
we have
 $|S(\hat{\vp}_{\KuI})-S(\vp_{\KuI})|\le 
-1/2\log1/2-1/2 \log1/2=\log 2$,
  and similarly $|S({\hat{\vp}}_{\I})-S(\vp_{\I})|\le \log 2$
 and 
  $|S({\hat{\vp}}_{ \K})-S(\vp_{\K})|\le \log 2$.
 Hence
$\vert  S(\vp_\I) -  S(\vp_\K)\vert-S(\vp_{\KuI})$
 is at most $3\log2$.
However, we do not know its    possible maximal
 value. 
(The violation of  the  triangle inequality
   for our concrete    model  
  considered in Ref. \citen{ENTA}
ranges    from $0$ up to $\log 2$.) 

\end{document}